# Testing the Brain Wave Hypothesis

Robert Worden

Theoretical Neurobiology Group, University College London, London, United Kingdom

rpworden@me.com

Abstract:

It has been proposed that there is a wave excitation in animal brains, whose function is to represent three-dimensional space around the animal as a working spatial memory. After surveying the evidence supporting the hypothesis, I discuss ways in which it can be tested.

There are many ways to investigate it, theoretically and experimentally. They include connectome studies, computational modelling, experimental neuroscience, genomics and proteomics, studies of animal behaviour, and biophysics.

If the wave exists, there is a compelling case to identify it as the source of consciousness. This would advance our understanding of one of the greatest scientific challenges of all time, while changing our view of the human mind.

# 1. Introduction

It has been proposed that there is a wave excitation in the brain, whose function is to represent the three-dimensional space around any animal, as a working spatial memory. The proposal is described in [Worden 2020a, 2024b, 2024c].

Wave-like electromagnetic fields exist in the brain, and have been proposed for several roles [Pinotsis, Fridman & Miller 2023, McFadden 2002]. The wave discussed here has a different role, to act as a working memory for 3-D spatial information, for periods as long as fractions of a second. Electromagnetic fields do not store information for those periods, so the proposed wave is probably not just electromagnetic. Although electromagnetism may be involved, something else is required.

[Worden 2024d] estimated the probability of the wave hypothesis from existing evidence, and found it to be greater than 0.4; not negligibly small, but equally, not confirmed. In view of the potential importance of the wave hypothesis, it is a priority to confirm it or refute it.

This paper describes possible further investigations to test the wave hypothesis, in an initial survey of work that can be done – ranging from direct tests that can be made in the short term, to open-ended avenues for research.

A summary table of the proposed tests of the wave hypothesis follows:

| Sub-sect. | Investigation |
|---|---|
| 3.1 | Thalamus connectome tests |
| 3.2 | Insect central body connectome tests |
| 3.3 | Detailed thalamic neuroanatomy |
| 3.4 | Detailed neuroanatomy of the insect central body |
| 4.1 | Multiple Species comparisons |
| 4.2 | Single-celled animal cognition |
| 5.1 | Targeted genomic and proteomic searches |
| 6.1 | Neural computational models of spatial cognition |
| 6.2 | Wave-neural computational models of spatial cognition |
| 6.3 | Computational models of sense data steering and binding |
| 7.1 | Insect behavioral studies |
| 7.2 | Small animal behavioral studies |
| 8.1 | Data from phenomenal consciousness |
| 9.3 | Quantum coherence in biological matter |
| 9.4 | Nuclear and electron spin effects |
| 9.5 | Bose-Einstein Condensates |
| 9.6 | Sustaining the wave, and coupling neurons to the wave |
| 10 | Direct detection of the wave |

The table gives an impression of the wide range of disciplines that can contribute..

If the wave hypothesis is correct, there is a compelling case to identify the wave as the source of consciousness [Worden 2024e]. The function of the wave has a good fit to the spatial nature of consciousness, as we experience it in every moment of the day. Wave-based consciousness can solve many of the problems which have beset neural theories of consciousness. It could advance our understanding of one of the greatest scientific challenges of all time. For this reason alone, it is important to confirm or refute the wave hypothesis.

# 2. Current Status of the Wave Hypothesis

Generally, and as in the Bayesian philosophy of science [Sprenger & Hartmann 2019] any hypothesis is to be assessed by the Bayesian posterior probability that it is correct, in the light of all the evidence. In [Worden 2024d] I have applied this to the hypothesis of a wave in the brain], and found the probability of the wave to be greater than 0.4. I request commentary on that assessment, and summarise the main considerations here.

I first note two possible reasons **not** to believe the hypothesis – reasons which may occur to readers. These reasons are:

- We have studied brains for many years using a classical neuron model; we know what neurons in the brain do; they need not couple to any exotic wave in the brain.
- If there was a wave in the brain, we would have detected it by now.

I suggest that neither reason has much force.

We have studied brains for many years; and they are typically modelled with a simple model of the neuron [McCulloch & Pitts 1943, Hebb 1949] which under-utilises the capabilities of the eukaryotic cell. This classical neural model has not yet been so successful and comprehensive that (as with classical physics at the end of the nineteenth century) we can look forward to some modest tidying-up for the textbooks. There is still a huge amount still to be understood about brains. We should not be too attached to the assumption that the classical neural synaptic model does all computation in the brain.

Should we have detected any wave in the brain? We know that neurons can couple to waves at very low intensities (to light, down to the one-photon level), and we would expect that any wave in the brain used for working memory would have evolved to have the lowest possible intensity, to minimize its energy consumption. There are reasons (described in section 9) to expect that the wave is not simply an electromagnetic wave or field. We have not yet detected a wave in the brain, simply because we have not known how to look for it – and when we do, it will probably be hard to detect. In this respect, it is like the neutrino [Pauli 1930; Cowan et al 1956], or dark matter.

So these negative reasons have little force. They do not imply a very small probability for the wave hypothesis, or do much to counter the positive reasons for believing it. These reasons have been described in previous papers [Worden 2020a, 2024c] and are assessed in [Worden 2024d]. See also the discussion of the hypothesis at https://www.youtube.com/live/zqOcywx40n8 .

There are three main lines of evidence for a wave excitation in the brain, representing 3-D space:

1. **The Mammalian Thalamus**: The round shape of the thalamus is well suited to hold a wave, and is remarkably conserved across all mammal species. The thalamus is well connected to hold spatial information, having links to most sense data and to the cortical regions that use it. There is also a 'smoking gun' in that without assuming a wave, the neuroanatomy of the thalamus does not make sense energetically.
2. **The Central Body of the Insect Brain**: The near-round shape of the insect central body is well suited to hold a wave, and is remarkably conserved across all species. The central body is neurally well connected to hold spatial information. The number of neurons in the insect brain is insufficient to support 3-D spatial cognition, yet insects do it well.
3. **Computational Adequacy**: In spite of the central role of 3-D spatial cognition in the brain, there are no working neural computational models of it. (Existing models focus on learning shapes, rather than the more fundamental problem of representing 3-D space; and they generally ignore the issue of storage imprecision from neural spike trains). Neural storage errors for 3-D positions are too large; neurons are too imprecise and too slow. Wave storage of positions can solve this problem.

These three lines of evidence are described in the sections which follow, describing how they can be investigated further.

To summarise the current balance of evidence for the hypothesis: there is more evidence for the wave hypothesis than there is evidence against it. The evidence for the wave is quite strong; there are striking facts of neuroscience which cannot be understood without assuming a wave. This evidence is assessed in [Worden 2024d], leading to an estimate of 0.5 for the probability that the wave hypothesis is correct. This estimate is robust, in that it assumes only a very small prior probability for the wave, but the positive evidence is sufficient to counter that. Equally, it does not give a probability as high as 0.95, which would be a minimal 'one standard deviation' criterion to believe the hypothesis.

The aim of this paper is to rectify that – to discuss how to obtain the evidence to confirm or refute the wave hypothesis.

The rest of the paper describes lines of research to test the wave hypothesis. Possibly the first thing to do – the quickest way to make progress – is to 'stress-test' the three lines of evidence noted above, by looking more closely at them. Other tests follow after that.

## 3. Neural Anatomy and Physiology

### 3.1 Thalamus Connectome Tests

The first evidence for a wave in the mammalian thalamus is the regular round shape of the thalamus, which is highly conserved in all mammalian species. A round shape is significant because it well suited to hold a wave, with wave vectors in all three dimensions – representing things distributed in three dimensions around the animal. The thalamus in a few mammalian species is shown below:

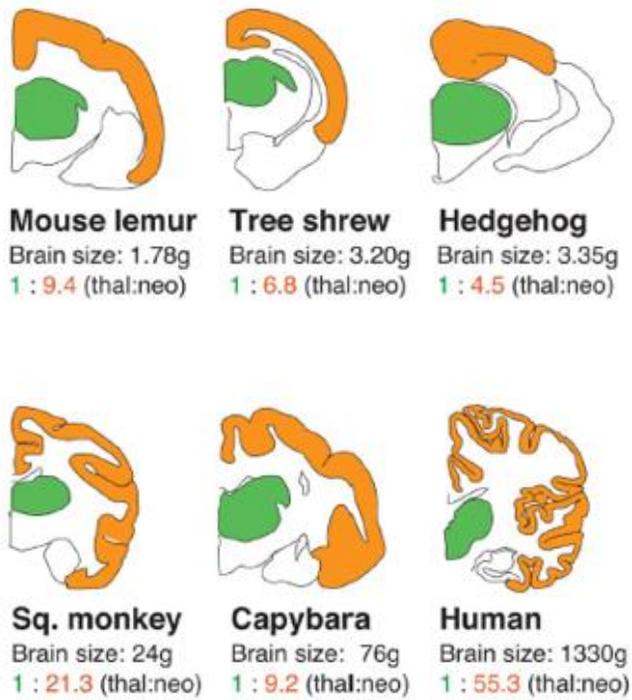

*Figure 1: the shape of the thalamus (green) compared to the cortex (orange) in a few mammalian species, from Haliey & Krubitzer 2019*

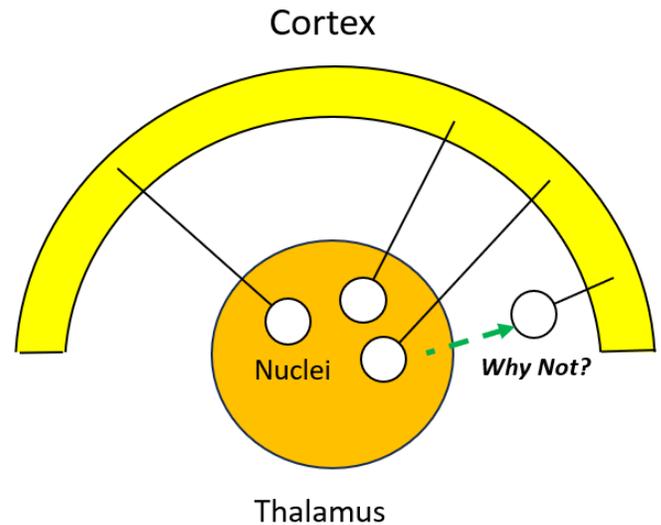

*Figure 2: If there is no wave in the thalamus, thalamic nuclei can migrate outward towards the cortex, saving energy and keeping the same neural computational capability.*

As well as its round shape, the thalamus has all the right neural connections – to incoming sense data, and to other parts of the brain – to serve as a spatial working memory, as is proposed in the wave hypothesis.

The round shape of the thalamus in many species is suggestive of a wave, but it is not the most definitive evidence. There is other anatomical evidence which is hard to account for in a purely neural synaptic theory of brain function – a smoking gun for a wave. I next describe the smoking gun argument as it has been developed so far, before describing how it can be made more precise by using connectome data.

The reasoning [Worden 2010] concerns brain energy consumption, caused by the length of myelinated axons. All thalamic nuclei have reciprocal connections to many parts of cortex; but the cross-connections between different thalamic nuclei are weak or non-existent [Sherman & Guillery 2006]. Individual thalamic nuclei could migrate outwards towards the cortex, separating from each other, and reducing the length of their thalamo-cortical and cortico-thalamic axons. By this change, net axon length would be reduced, so brain energy consumption would be reduced, while keeping the same neural synaptic connectivity - so keeping the same neural computational ability, in the classical neural model of computation. This possible change in thalamic anatomy is illustrated in figure 2 below.

According to this reasoning, we would expect the thalamus not to be a near-round central part of the brain, but to be fragmented , or at least to be irregular and species-specific – like other parts of the brain such as the hippocampus. Its compact regular shape does not make sense.

A way to understand the regular, compact shape of the thalamus in all mammalian species is to assume that thalamic nuclei all need to stay close to one another, to be immersed in the same wave. It is hard to think of another account of the neuroanatomy of the thalamus; its roundness specifically suits a wave in three dimensions. This is the smoking gun for the wave[1].

The possible savings in brain energy consumption from an 'exploding thalamus' were approximately calculated in [Worden 2010]. This calculation used approximate neuroanatomy in *Homo Sapiens*, and concluded that savings in energy consumption would result from migrating thalamic nuclei towards the cortex.

Those calculations were approximate, so they leave reason to doubt the conclusion. It is possible to make the same calculations much more precisely, using connectome data (for instance from the human connectome project, at https://www.humanconnectome.org/), with automated computation of the energy consumption in varied anatomical configurations. The program required to do this need not be complex. It only needs to explore perturbing thalamic neuro-anatomy in various ways (which are anatomically possible) - and for each perturbation, re-compute the brain energy consumption from the net axon length. This is the kind of repetitive calculation that computers do well.

---

[1] The paradox of the thalamus staying together, and not disintegrating, is like the paradox of galaxies not disintegrating, noted by Zwicky in the 1930s, which led to the hypothesis of dark matter. The wave is like the dark matter of the brain.

The program can compute not only finite perturbations to thalamic anatomy, but can also compute in the limit of small perturbations, to test whether the shape of the thalamus is stable against small perturbations in any direction. If it is not stable (if the axonal energy consumption has a non-zero spatial gradient), the program can find a minimum of energy consumption, by gradient descent minimization.

Calculation of the gradient of the energy consumption is particularly simple using a 'virtual work' analysis, like that used for mechanical problems. Each axon is modelled as having a 'tension' (coming from the energy saved by reducing its length). The tensions of all axons connecting to a specific thalamic nucleus are added as forces (by vector addition), to find a net force on the nucleus. If the net force is non-zero, then energy can be saved by the nucleus moving in the direction of the net force.

Given access to connectome data, this appears to be a straightforward research project.

A possible alternative to the wave account is to assume that there are subtle neural timing requirements, which determine thalamic neuroanatomy. However, if precise timing is required, the existing timing can be preserved by shortening the axons, and reducing their level of myelination to preserve the timing – again while reducing energy consumption.

The second barrel of the smoking gun concerns the Thalamic Reticular Nucleus (TRN), and was analysed in [Worden 2014]. The anatomy of the TRN is very distinctive; it is a thin shell surrounding the dorsal thalamus[2]. In a pure neural synaptic model of computation, in energy terms this anatomy does not make sense. Net axon length could be saved (and energy consumption reduced) if the TRN moved partially inside the body of the thalamus, reducing the TRN's area, rather than stretching it round the outside. The energy computations of this possible movement in [Worden 2014] were approximate and theoretical. They can now be re-done with greater precision using connectome data. This is another fairly straightforward computational project.

The anatomy of the TRN – a thin shell around the body of the thalamus – is suggestive of a wave. For instance, it might suggest that the TRN holds transmitters for the wave, while the body of the thalamus holds the receivers; or that the TRN somehow maintains boundary conditions for the wave. It may not be as simple as this; but the distinctive shell-like form of the TRN is a clue.

### 3.2 Insect Central Body Connectome Tests

The shape of the insect central body is approximately round, and is highly conserved across all insect species[Strausfeld 2011; Heinze et al. 2023] and is suggestive of a space holding a wave [Worden 2024c]. The central body of a bee brain is shown in figure 3 below.

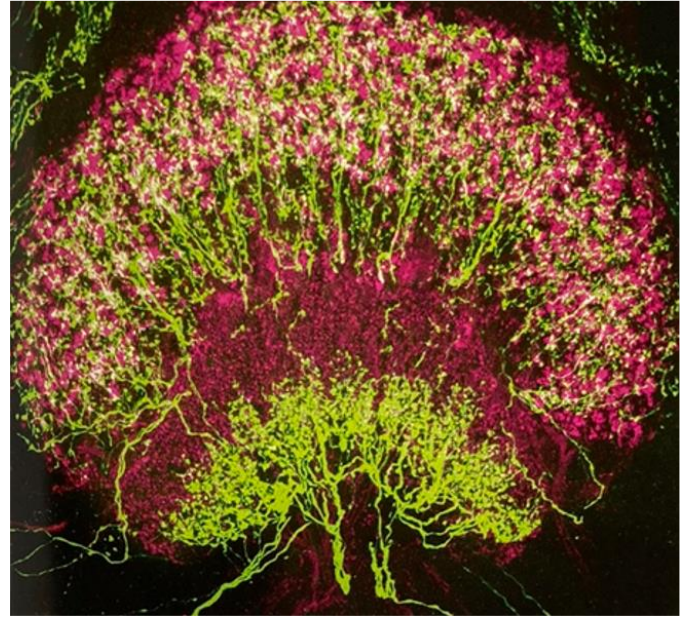

*Figure 3: central body of a bee brain, from [Strausfeld 2011]*

The central body does not have separate nuclei like the thalamus; rather, many of the neurons that innervate the central body appear to have synapses distributed throughout its volume. This in itself is suggestive of a wave. Having the synapses of one neuron distributed throughout a volume (assuming those synapses act as transmitters or receivers) allows the neuron to be selective in the wave vectors it couples to. Having selectivity in all three dimensions is the main reason for expecting a nearly round shape, with comparable extension in all three dimensions.

Because the central body does not have separate nuclei, it is not easy to devise a simple perturbation to its anatomy (like separating the thalamic nuclei) which would lead to energy savings. Nevertheless, there may be ways of perturbing the anatomy, while not changing neural synaptic connections, which would reduce net axon and dendrite length. So a computer program, using connectome data, could search for perturbations to the neuroanatomy of the central body which could reduce axon length and so save energy, while preserving connectivity. It could search for incremental perturbations, using the analogy to mechanical forces. Possible perturbations include making the central body elongated or flat, or changes to the distribution of synapses. If the program finds any direction of anatomical change with a non-zero energy derivative, it can follow that change by gradient descent minimization, to a minimum of energy consumption. This would raise the question: why is the central body nearly round, rather than that other shape which would reduce energy consumption?

---
[2] I know of no other similar structure to the TRN in the brain; others may know of them.

This is also a fairly simple computational project, given insect connectome data. There is a complete connectome for the fruit-fly larva [Winding et al 2023]; however, larvae have much less need than adult insects for fast spatial cognition, so looking at the central body of a larva may not be conclusive.

### 3.3 Detailed Thalamic Neuroanatomy

If the thalamus holds a wave excitation, and if neurons couple to the wave as transmitters and receivers, then we would expect to see clues to that in its detailed neuroanatomy. Thalamic neuroanatomy is distinctive, well studied, and in many ways unlike other parts of the brain [Sherman & Guillery 2006; Jones 2007; Halassa & Sherman 2019]. However, it is not easy to say what detailed features might be correlated with a wave. Some possibilities are:

- Compound synapses and glomeruli: If some neurons couple selectively to a wave, then other neurons may control or modulate the ways in which they couple. The locus of this control could be compound synapses or glomeruli, where several neurons come together [Spacek & Lieberman 1974].
- Laminae in the LGN: These intriguing structures vary across species (e.g. in the number of laminae). The LGN relays visual data to the cortex. Could the laminae be linked to waves representing different ranges of depth? Can their spacing and orientation be related to a wave?
- Neural maps: There are some two-dimensional neural maps in the thalamus (in the distribution of cell nuclei?); do they have any connection with a wave – given that the wave may couple to neurons via transducers distributed across their dendrites and axons, rather than at their nuclei?
- Sherman & Guillery [2006] have identified a distinction between 'driver' and 'modulator' pathways in the thalamus; this distinction pervades all nuclei and many aspects of the thalamus. Nobody has attempted to relate this distinction to the wave hypothesis; looking for a relation is an open area of research.
- Jones [2007] identified a 'core-matrix' distinction between calbindin- and parvalbumin- immunoreactive neurons. Could this distinction relate to the wave hypothesis?

Neural geometry will surely come into these questions. For instance, for a neuron to be highly spatially selective, its transmitter or receptor units (possibly in its synapses) must be widely spaced, far from the cell nucleus.

This research topic is currently open-ended and undefined, with few specific clues about where to look. It is both a search for transducers that couple neurons to the wave, and a search for the mechanism that sustains the wave over time.

### 3.4 Neuroanatomy of the Insect Central Body

Just as for the mammalian thalamus, if the insect central body is the site of a wave excitation, and if neurons couple to the wave as transmitters and receptors, there may be clues to this in the detailed neuroanatomy [Strausfeld 2011; Heinze et al. 2023]. Again, however, it is hard to know what to look for. It is an open-ended research topic.

It may be worth looking for analogies between structures in the insect central body, and structures in the thalamus. This might narrow the search.

A key difference between insects and mammals is the number of neurons available in the brain. This makes it harder to defend a classical neural model of insect spatial cognition.

## 4. Multiple-Species Brain Studies

The first two lines of evidence for a wave in the brain come from opposite ends of the animal evolutionary tree – insects and mammals. If the wave hypothesis is correct, wave spatial memory evolved before these two lines diverged; so it should also be found in all other species which have the same common ancestor – approximately 500 million years ago.

### 4.1 Brains of Larger Animals

There may be neuro-anatomical evidence for a wave in the brain in all animals with limbs and complex sense data – including birds, reptiles, amphibians, fish, arthropods, and cephalopods. This group includes all species with capable eyes.

For this wide range of species, the wave hypothesis makes a distinctive prediction, which (if wrong) is falsifiable. In all those species, it predicts that there will be a part of the brain which plays the same wave-related role as the mammalian thalamus, as indicated by the following properties:

- Approximately round shape, suitable to hold waves in all three directions
- Size large enough to hold a large number of waves, of some minimum wavelength
- Near-central position in the brain, to communicate with many brain regions

- Innervated by sense data of all modalities (except possibly olfaction)[3]
- Having the neural connections required for multi-sensory integration

All vertebrate species have a thalamus. This in itself is evidence supporting the wave hypothesis. [Lynn, Schneider and Bruce 2015], reviewing the avian thalamus, state that: '*a dorsal thalamic region is recognized in all vertebrates and believed to be homologous*', while [Mueller 2012] states that: '*research on the thalamus and related structures in the zebrafish diencephalon identifies an increasing number of both neurological structures and ontogenetic processes as evolutionary conserved between teleosts and mammals*'. So there are indications that in all vertebrates, the thalamus may play a wave-related role. There is scope for more detailed comparisons across species, to confirm or refute this.

Particularly interesting is the case of cephalopods, because the octopus is well studied, and has advanced spatial cognition. In a review of the cephalopod brain, [Shigeno, Andrews, Ponte and Fiorito, 2018] suggest that '*in the cephalopods, dorsal basal- and sub-vertical lobes could be considered as candidates for analogs to the vertebrate thalamus*'. While they did not have a wave-hosting role in mind, this could be a starting point to look for a possible wave excitation representing space in the cephalopod brain.

Insect brains and other arthropod brains such as crustaceans [Strausfeld 2011] can be included in these cross-species comparisons.

In summary, there is the potential for collaboration between researchers in the brains of many species, to further probe the hypothesis of a wave excitation in animal brains, representing the 3-D space, across all animal species.

### 4.2 Single-Celled Animal Cognition

Some single-celled animals have complex behaviour, and they appear to use spatial signals to control it. The single-celled *Erypthosidinium* (some 50 μm in diameter) has an ocelloid eye, shown in figure 2. It hunts and eats crustacean eggs. If all brains require neurons, this is a mystery. How does *Erypthosidinium* use the information from its eye, if it has no brain?

Other single-celled animals in the group *Warnowicii* have similar ocelloid eyes and behaviour, with different body plans (including other appendages), illustrated in [Gomez 2017], which is an extensive behavioral study of *Erypthosidinium*.

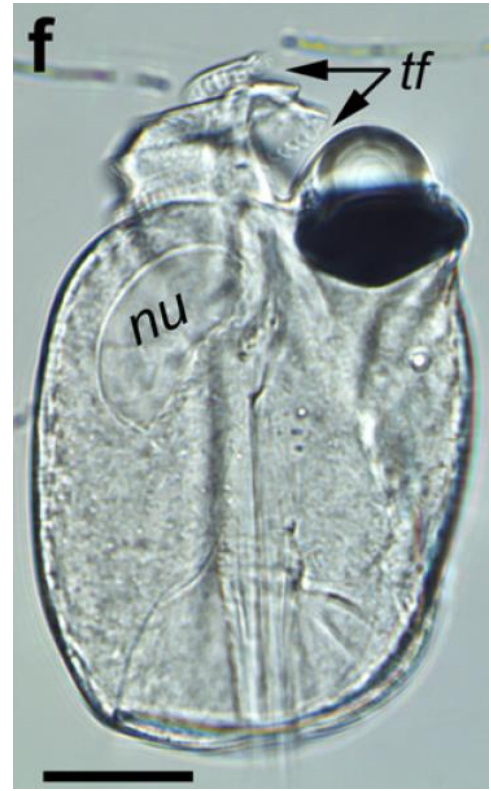

*Figure 4: The body of Erypthosidinium, showing the lens of the ocelloid eye (round shape at top right), the retinal body (dark mass below the lens), the cell nucleus (nu) and the transverse flagellae (tf). The central downward line holds the piston, which can be rapidly ejected to distances up to 700μm, to propel the animal, attach to objects or catch food. Figure from [Gomez 2017]. The scale bar is 20 μm.*

If the brains of higher animals contain a wave excitation that can persist for short periods, then the capability to sustain a wave must be a capability of their cells. It is then reasonable to suppose that single-celled animals use the same capability, if there is a need it.

A wave in a single cell could sustain some directional information (such as light or sound) for short periods – after which time, it can be compared with later directional information, for instance to get a fix on the range of some light source (as a simple way of inferring depth as Structure From Motion, SFM) [Murray er al 2003]. This suggestion is merely illustrative, to point out that a single cell is capable of more complex adaptive behaviour if it can sustain information for short times in a wave excitation. How it does so is not clear; but just because there are no neurons, we should not underestimate the capabilities of the eukaryotic cell, to carry out useful computation.

More generally, the eye-like structures in these cells call out for an explanation. A persistent wave is one of many possible explanations. Behavioral studies and micro-anatomical studies of single-celled animals will lead to further hypotheses –if not to a wave.

---

[3] Because smells diffuse slowly, they are not as useful for rapid precise spatial cognition as sense date of other modalities. That may be the reason why olfactory sense data do not pass through the mammalian thalamus on their way to cortex.

In section 9, when considering the physical substrate of the wave, an important topic is quantum coherence – which, when it occurs in living matter, evidently occurs within single cells. In this regard, from cellular energy considerations [Fields & Levin 2021] conclude that '*cellular information processing must employ quantum coherence as a resource for reversibility'*. Quantum coherence may play a role in single-celled cognition in *Warnowicii*.

## 5. Genomic and Proteomic Studies

If the wave in the brain exists in both insects and mammals, then its evolution pre-dates their divergence from a common ancestor, and we would expect to find it in all species with capable spatial cognition descended from that ancestor. It may depend on the capabilities of single cells to emit and detect the wave; so we might expect to find the same or similar single-cell capabilities across all those species. It is less likely (but possible) that analogous single-cell capabilities would have evolved separately in many species, because the need for it arose in the Cambrian era, before those species diverged.

### 5.1 Targeted Genomic and Proteomic Search

We might expect similar proteins to underly the wave capability, across a wide range of species. We can look for the proteins, and for the genes needed to make them – for instance, in the human protein atlas [Sjöstedt et al. 2020] at https://www.proteinatlas.org/humanproteome/brain/thalamus . The search can be constrained because we expect the genes to be expressed only in quite restricted regions of the brains – in neurons which originate in, or innervate, some near-spherical, near-central region of the brain, of which the insect central body and the thalamus are two examples. Similar genes and proteins might be seen across many species, in similar regions.

This defines some quite specific genomic and proteomic studies, which could help to confirm or refute the wave hypothesis.

Another way to narrow the search is to focus on proteins that anesthetic compounds bind to, since anesthetics interrupt spatial cognition. There is an indication that anesthetics probably act directly on the wave and the cells immersed in it, rather than on other neurons that define the form of the wave. This clue is that when going into anesthesia, people do not have near-death experiences; they just pass out. In a near-death experience (or in death), the neurons maintaining the form of the wave may fail before the wave itself fails, so that consciousness in the wave persists for a short time after the neurons fail; whereas under anesthetics, the wave fails first. This suggests that anesthetics act directly on the wave.

If some candidate proteins are identified though any of these routes, their properties may give clues about the biophysics of the wave.

## 6. Building Computational Models

This research area relates to the third main line of evidence for the wave in the brain, and is a way to stress-test that evidence. The existing evidence is the difficulty of building a computational model of spatial cognition using only classical neural computation – and the possibility that wave storage of 3-D positions could resolve the difficulty.

Neural storage of 3-D spatial data appears to be too imprecise, and too slow, to do 3-D spatial computations. Something else seems to be needed, and that something could be a wave.

Neural computational models of the brain (such as Wallis & Rolls 1997, Rolls & Deco 2022, Hawkins et al 2017, Reisenhuber & Poggio 1999]) almost always model the inputs and outputs of a neuron as real variables, using the high precision of modern floating-point arithmetic – because that precision is available. It is very uncommon to go to the extra length of modelling neuron outputs as spike trains, which they really are. As a consequence, the issue of the imprecision of spike trains is rarely addressed. The problem is shown below.

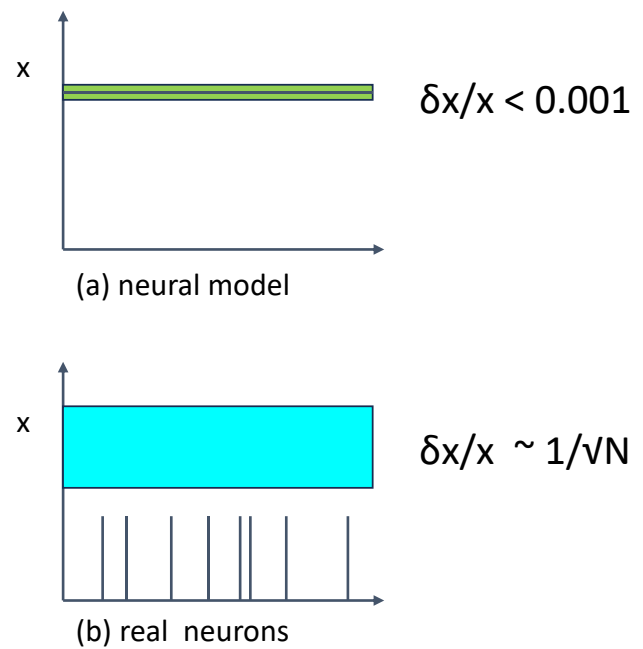

*Figure 6: (a) in nearly all neural models of the brain, the inputs and outputs of neurons are modelled as real variables, with high precision. In representing any real quantity x, the error δx is very small indeed. (b) real neurons represent a real quantity x by a spike train, which*

*does not give high precision in a short time interval; the relative error δx/x in a sub-second interval may be √10 = 30% or more.*

Spatial cognition requires both speed and precision. It appears that neuron spike trains are not capable of giving the required speed and precision – by quite a large margin (by a factor of the order of 30 – see below).

There is another reason why this problem has not received the attention it deserves. Most neural computational models of the brain (like those mentioned above) focus on the issue of learning, as a measure of their performance. In doing so, they do not address issues of speed and precision.

For the purposes of this paper, learning is not the primary issue. The proposed purpose of the wave is to hold a model of 3-D space around the animal, analogous to a map. In this context, learning is learning how to classify the landmarks on the map. There is an earlier and more basic problem, of drawing the map and making it geometrically faithful to reality, treating the landmarks just as 'things', and knowing where they are, regardless of their classifications.

Many newborn animals (such as a newborn horse or deer) do this before they have had time to learn anything. For other animals (for instance, those living in deserts) the problem is not to classify the shape of some newly encountered rock; it is simply to understand its irregular shape, so as to tread on it without falling over.

So while models of spatial learning are of interest, they do not directly address the issues of this paper. Learning is a discrete categorisation problem, and there are neural models which can do it. Building the 3-D spatial map is a geometric problem of continuous variables, where precision matters. That is why a wave may be particularly suited to do it.

A Bayesian computational model of 3-D spatial cognition [Worden 2024b] has been built at Marr's [1982] Level 2, of data structures and algorithms. This model builds the 3-D map; it does not learn to classify the landmarks. In this, the model performs well, as animals do. It requires precision in storing positions of the order of 1%, with response times less than 0.1 second (as is needed to update the spatial model in small animals). If positions are stored as neural firing rates, this precision and speed cannot be achieved – there is too much random neural noise.

If a model at Marr's Level 2 requires those levels of speed and precision, then it is likely that any neural implementation of that model (or a similar model) would require the same precision and speed. No such neural model has yet been built; but wave storage of positions might give the required performance. This motivates the hypothesis that there is a wave in the brain, used to store 3-D positions.

This leads to a twin-track test of the wave hypothesis: first, to try to build a working neural model of spatial cognition, to see if the difficulties of neural speed and precision can somehow be solved (in which case the wave may not be needed); and second, to build a hybrid implementation, with neural computation and wave storage of positions. Both models need to be evaluated for their scaling and performance, compared to animal performance. These studies are described in the following sub-sections.

Richard Feynman once wrote: "If I cannot build it, I do not understand it." To understand 3-D spatial cognition, we need to build a model of it, paying attention to issues of scaling and performance.

### 6.1 Neural Models of Spatial Cognition

It is now known that much animal cognition is Bayesian. Animals use Bayes' theorem (or good approximations to it) to infer the most likely internal model of the state of the world, given their sense data [Knill & Pouget 2004]. Worden [2024a] has shown that this applies to 3-D spatial cognition, using sense data from vision, proprioception, touch and other senses. If animals build a Bayesian internal model of the 3-D space around them, the Bayesian model gives greater lifetime fitness than any other form of spatial cognition. So a Bayesian spatial model is the form of spatial cognition that evolution converges towards (and there has been huge and sustained selection pressure to build the best possible spatial model). This motivates building Bayesian neural models of spatial cognition. The optimal Bayesian internal model of local 3-D space can be computed in modern digital computers (although probably not computed neurally in animal brains), so it can serve as a 'gold standard' for comparison with any neural implementation or data about animal spatial cognition.

The Free Energy Principle (FEP) [Friston 2003; Friston, Kilner & Harrison 2006; Friston 2010] is a Bayesian model of cognition, which has a precise computational model with a neural implementation (its neural process theory) [Parr, Pezzulo & Friston 2022; Smith, Friston & Whyte 2022]. It has been successfully compared with animal cognition in many domains. Active Vision [Parr, Sajid, da Costa, Mirza & Friston 2021] is the FEP framework for modelling 3-D spatial cognition. Active Vision is a suitable framework for building a neural model of animal 3-D spatial cognition.

In the neural process theory of Active Vision [Smith, Friston & Whyte 2022], quantities such as the components of spatial 3-vectors are represented by idealised abstract neurons, whose inputs and outputs are real variables with high precision. To address the question of neural imprecision, these precise quantities need to be represented by stochastic neural firing rates; or they need to be 'seeded' with adjustable error levels to test the level of errors that can be tolerated.

If a neuron representing any quantity fires N times in a time interval, the expected random proportional errors in the quantity represented are of the order of $1/\sqrt{N}$. With typical neural firing rates of 5-50 pulses in a second, and a time interval of 0.2 seconds, the expected error rate is $1/\sqrt{10}$, or 30%. This is much larger than the error rate of 1% which [Worden 2024b] found to be necessary to build a working spatial model. I would expect that a direct application of Active Vision, with realistic neural error rates, will not make a viable 3-D spatial model – confirming the difficulty of a neural representation of space. Something else is required. Can it be done by neurons, or is a wave required?

One purely neural possibility is to use parallelism, with many neurons representing some quantity, to increase the effective firing rate and lower the errors. However, to reduce the error rate by a factor 30 requires 30*30 = 900 neurons in parallel, to represent each component of a position, possibly for large numbers of objects being tracked at any instant. For small animals such as insects, the required number of neurons would be prohibitive, and appears unlikely.

It is then necessary to consider more sophisticated neural representations of positions, which could give higher precision in short timescales. I do not know what these representations might be; but whatever they are, they would add complexity to the neural process model of computation. For instance, if a representation of space based on grid cells [Hafting et al 2005; Moser et al 2015] was proposed, as it has been in [Hawkins et al 2017], it would be necessary to check its scaling and performance properties; and there still remains the issue of how the neural computations are done.

Starting to address this problem in Active Vision would be a direct application of the well-established FEP formalism, and is a possible near-term project. It could start with abstract idealised neurons of very high precision, and then progressively turn up the error levels, to see what neural errors can be tolerated. Assuming that the errors from simple stochastic neural firing rates are too large, other options can be explored – extending the neural process model to make the required computations.

The challenge of neural imprecision can be addressed in other models, outside the Active Inference formalism. There are several models of neural spatial cognition, including large-scale models of three-dimensional spatial cognition [Rolls & Deco 2002; Wallis & Rolls 1997; Hawkins et al 2017; Riesenhuber & Poggio 1999 ]. As far as I know, these models all treat the inputs and outputs of neurons as real variables with very high precision – not as stochastic spike trains. It is hard enough to model spatial computing by neural networks, let alone to model the stochastic neural error rates. But that is what needs to be done, to understand the problem of neural imprecision. Computational models need to move fully down from Marr's Level 2, to the Level 3 of neural implementation – not just half way down.

Particularly problematic for a realistic neural implementation is the issue of **gradient descent.** Many neural models of the brain rely on minimization of some quantities, typically by gradient descent – following the gradient of a quantity iteratively downwards, until the gradient is zero, at the minimum. However, if neural errors are modelled, and if the gradient of a quantity is calculated naively – by computing the quantity at two nearby points and taking the difference – the proportional errors in the difference are very large, making gradient descent useless. Gradients of quantities need to be computed and represented separately from the quantities themselves. This is not simple. Some of the difficulties are:

- Generally, it is more complex to compute a gradient of a quantity, than it is to compute the quantity itself. More complex algorithms are needed.
- There needs to be mutual consistency between the algorithms – between the computation of a quantity, and the computation of its gradient.
- In spatial problems with three dimensions, there are three gradients for any quantity. Sometimes it is necessary to compute a 3x3 symmetric tensor of second derivatives[4] with respect to position (in the FEP, these are referred to as 'precisions' – unlike the use of the term precision in this paper).
- Any component of a gradient can be positive or negative. This means that a gradient cannot be represented directly by a single neural firing rate, which is always positive. Either a zero gradient must be represented by a non-zero firing rate (which is unattractive, implying large firing rates and large energy consumption in the 'resting' state of zero gradient), or the sign of a gradient must be represented separately from its magnitude. This makes any neural computing design more complex.

These are difficult issues, which have rarely been addressed in neural implementations taking account of stochastic neural noise and errors. The problem might be solved by some neural 'micro-architecture' or computing component, which is capable of giving high precision in short timescales, and of representing gradients. It is not clear that it can be done at all (the wave hypothesis is that this cannot be done); it may need many attempts to explore the possibilities.

Completing this project successfully is a hard computational challenge – not a simple project. But given the essential role of spatial cognition in any animal's survival, if there is no wave it is a high priority for neuroscience to show that some neural model of spatial cognition has sufficient precision to be viable. A theory of neural cognition which cannot handle spatial cognition is like a theory of planetary motion with no sun; or a theory of the atom with no nucleus.

---

[4] In principle, second derivatives need to be computed in gradient descent, to know how large a step to take in the direction of a first derivative, to find a minimum.

If you do not believe there is a wave in the brain, and prefer to work on classical neuroscience, this is a problem that will need to be solved.

Finally, it seems unlikely that 3-D spatial cognition will be solved by trying out large neural nets, until we find one that works – especially when neural stochastic noise and imprecision is modelled (as it is not in most neural net models). After forty years of neural nets, that does not look like a promising research agenda – especially for the small insect brain.

## 6.2 Wave-Neuron Computational Models

One solution to the problem of neural imprecision is to represent locations in space not by neural firing rates, but by a wave in the brain.

The essence of the wave hypothesis is this: There is some near-spherical region in the brain, of diameter D, which can hold waves of minimum wavelength λ. Neurons couple to the wave as transmitters and receivers, and the wave can sustain itself for times as long as a fraction of a second. It acts as a working memory for 3-D spatial positions of objects, in an allocentric frame of reference (a frame in which most objects do not move; persistence of the wave stores static object positions). A wave excitation with wave vector **k** represents an object at position[5] **r**, where **r**= α**k** and α is a constant. $|\mathbf{k}| = 2\pi/\lambda$.

The number of independent object locations which can be stored in the wave is of the order $(D/\lambda)^3$, which may be very large. The proportional error in any object position is (λ/D). This can easily be as small as 1%, giving high spatial precision.

When a neuron couples to the wave, it is assumed to do so through a large number of transducer units, which are widely distributed through the wave volume (to be sensitive to small changes in the wave vectors). Transducer units could be inside axons or dendrites, or might be anatomically visible, appearing like synapses or compound synapses. It is assumed that the transducers are sensitive to the phase of the wave[6], which enables a neuron to couple selectively to waves with wave vectors in the region near some wave vector **k**, representing objects near some spatial location **r**.

In an elaboration of the model, the tuning of some neuron could be altered by inputs from other neurons. This might be done in compound synapses or glomeruli, in which inputs from other neurons alter the phase dependence of individual transducers. This elaboration is relevant for spatial steering and binding.

Compared to the McCulloch-Pitts [1943] neuron, this may seem to be asking a lot of a neuron. However, the eukaryotic cell has vast capabilities. In some sense, all the capability of any animal is latent in its stem cells. Compared with the known capabilities of eukaryotic cells, the wave hypothesis is not asking much of neurons.

The wave serves as a working memory for spatial positions, with high capacity, high resolution, and fast response. What is the combined wave/neural architecture for computing the 3-D spatial model?

[Worden 2020] has described a cyclic 'aggregator' model of spatial cognition. In the aggregator model, a central wave memory (e.g. in the thalamus) has two-way connections to a large number of neural 'knowledge sources' (e.g. in cortical columns), in a hub-and-spoke configuration. Each neural knowledge source is concerned at any moment with only a small region of 3-D space (a small range of wave vectors), and with a restricted number of information types in that region. In each cycle of the aggregator (possibly at 40 Hz) the central aggregator broadcasts the latest state of the model selectively to the neural knowledge sources (each knowledge source has receptor neurons tuned to a small region of space, or wave vectors); and later in the same cycle, each neural knowledge source sends back an update to the centre, computed from its local knowledge.

The aggregator model aligns closely with a blackboard role for the thalamus [Erman et al 1980; Nii 1986; Mumford 1991; Llinas & Antony 1993; Lee & Mumford 2003; Worden Bennett & Neascu 2021] the global workspace in GNWT models of consciousness [Dehaene, Changeux and Naccache 2011], and with the tracking 3-D spatial model of [Worden 2024a]. The wave aggregator acts as blackboard, or workspace, for spatial information. The neural knowledge sources do a variety of computational tasks, such as edge detection, stereoscopy, multi-sensory integration, or spatially invariant pattern recognition. They each do this locally for a small region of space, so they are not required to have high spatial resolution. The central wave holds the 'big picture' global model of space with high spatial resolution, and the model is updated once in each aggregator cycle. A central computational function is to add (to aggregate) the negative log likelihoods (free energies) from the different knowledge sources for each small region of space – combining them into a single Bayesian maximum likelihood model of the whole space, incorporating both local and global constraints.

The aggregator model and the 3-D spatial tracking model have both been implemented at Marr's [1982] level 2, but not as neural implementations at Marr's Level 3. To do so, we would require neural implementations of the local knowledge sources – probably initially for only a restricted range of functions, such as stereoscopy. We also need to model the neural-wave coupling. The model could later be broadened to full multi-sensory integration.

---

[5] To represent objects at very large distances from the animal, **r** may be a near-projective transform of Euclidean space.

[6] The carrier frequency of the wave is an entirely open question.

Parts of this model have already been built, as separate computational models at Marr's Level 2. The four models which have been built are:

1. **Spatial Cognition in Bees and Bats**: [Worden 2024b] This model shows how animals build a 3-D Bayesian spatial model of the objects around them, from their changing sense data as they move (vision in bees, echo-location in bats), using Structure From Motion (SFM), requiring a spatial working memory (to be held in the wave). It is likely that an FEP model of the same computation would give similar results; they both find the Bayesian maximum likelihood .
2. **The Aggregator Model of Spatial Cognition**: [Worden 2020a] This model shows how different cortical knowledge sources (such as edge detection, stereopsis, sound location, proprioception, or object recognition) collaborate through a star connectivity to a central aggregator (such as a wave in the thalamus), to build a single Bayesian maximum likelihood model of space, binding together sense data of different modalities. It is in effect an extension of the visual/sonar model of (1), to multi-sensory integration.
3. **Projective Transform Model**: This model shows how a projective transform of space draws in objects from very large distances to smaller distances, so that all perceived objects can be represented in the same wave with bounded wave vectors. A projective transform preserves straight lines and planes.
4. **3-D Fourier Transform Model**: This model shows how a conscious experience of objects in 3-D space arises as a 3-D spatial Fourier transform of the amplitude of a wave excitation. Relating to the aggregator model, each cortical knowledge source would be spatially steered to selectively sample a small volume of the Fourier transform.

Models (3) and (4) are just programs which compute and display a well-known piece of mathematics. Combining and extending these models is not trivial, but it gives a starting point.

While it might be possible to simulate the physics of the wave, this is probably not necessary in the first instance. It will be enough at first to simulate the memory properties of the wave – including its spatial precision, capacity, and spatially selective retrieval.

This is an ambitious computational modelling programme. Its strategy is the same as that of the previous section: take a computational model which is understood at Marr's Level 2, and progressively move it down to the neural implementation level 3, while checking its scaling and performance in the presence of neural stochastic errors. Hopefully, by using the wave for working spatial memory, it will not run into any brick walls of precision and speed; and it may clarify the required forms of coupling between the wave and neurons. It is perhaps easy to see how such a research program would start – harder to know where it would end.

### 6.3 Models of Steering and Binding

Brains are required to bind together sense data of different modalities coming from the same spatial location, and to route it to specialized modules in the cortex which recognize objects and predict their behaviour. Signal binding is a long-standing problem in neuroscience [Treisman & Gelade 1980; Treisman 1998; Feldman 2013]. Although binding by synchrony [von Der Malsburg 1995] has been proposed as a solution, synchrony is a weak computational mechanism [Shadlen & Movshon 1999] and has not been built into working, scalable models.

For spatially invariant pattern recognition, and for multi-sensory integration, some form of selective signal routing (or steering) is required [Olshausen, Andersen & Van Essen 1993,1995]. The natural basis for selectivity is the spatial origin of sense data, because sense data from the same location comes from the same physical object. So signal steering and binding may be linked to an animal's 3-D model of local space.

In a purely neural cognitive architecture, it is possible to devise signal steering schemes of a fan-in/fan out variety, with selective switchable synapses [e.g. Wallis & Rolls 1997]. However, there are many open questions about how synapses are selectively switched on and off, and how these architectures scale up to meet real animal requirements – the number of neurons and switchable synapses required could be prohibitive.

If a wave is used to store the internal model of local 3-D space, the wave has the potential to carry out a spatial steering function, based on the 3-D locations of sense data sources. This can be done using tunable receptor units on neurons, so that the range of **k**-vectors which a neuron is sensitive to can be tuned by some modulator input – almost like a steerable searchlight of attention for that neuron [Worden, Bennet & Neascu 2022]. Many details of this remain to be defined and explored, but it has the potential to give better scaling and performance than a purely neural switching architecture.

This suggests that the twin-track research agenda of the previous two sub-sections – trying to build a working computational model of 3-D spatial cognition, by neurons with a wave, and without it – should be extended to address the problem of signal routing and binding. If a purely neural solution is adequate and scales well, then perhaps a wave is not needed. On the other hand, if a wave solution is clearly better than a pure neural solution, that supports the wave hypothesis.

## 7. Experimental Studies of Animal Cognition

Brains evolve towards Bayesian cognition, and the optimal Bayesian model of 3-D space which can be built from sense data is a precise model [Worden 2024b].

This raises the empirical question: just how good is an animal's internal model of the 3-D space around it? Is it nearly as good as the optimal Bayesian model[7], or is it significantly inferior? This question can be addressed by empirical studies of animal behaviour, designed to study the performance (e.g. the precision, capacity, and response times) of their internal 3-D spatial models.

For these experiments, it is possible to compute the optimal Bayesian model of 3-D space from measurable properties of the animal's sense data, such as its visual acuity. This optimal spatial model gives a yardstick, against which animal performance can be measured – does it come close to the best possible, or not?

### 7.1 Insect Behavioral Studies

Insect brains are of interest because they contain so few neurons. In the absence of wave storage, it is hard to see how insects could build an internal 3-D model of space, with precision comparable to their vision – let alone update it the very short timescales of insect behaviour. It is not possible invoke large numbers of neurons to somehow fix the problem. So the gap between 'neurons only' and 'neurons + wave' should be easiest to detect in insects. Also, insects may easier to rear, train and test than some other animals. Bees and insects such as *drosophila* can be trained [Chittka 2022].

What experiments can be done to probe the quality of the insects' 3-D model of space – to test behaviour that does not depend only on raw visual data, but which requires a 3-D internal model of space to guide it? I only offer initial suggestions. There are many possibilities for experimental designs, for instance using virtual reality to 'spoof' the insect's perception.

Two important spatial variables, which depend on an insect's 3-D model of space, are:

1. The range from the insect to some object.
2. Detection of motion while moving, possibly where only the depth of the moving object changes (so that direct detection of motion from the visual field is harder)

For instance, bees could be trained to gather nectar preferentially from flowers which move in a depth-only manner, to measure their performance as the amount of movement is changed. Or there might be a task which depends on detecting collinearity (or the lack of it) in 3-D space.

### 7.2 Small Animal Behavioral Studies

There is a wide range of potential animals to test – such as small mammals, birds, or fish, and the issues of experimental design for these species are similar to the issues for insects – to test some task which measures the quality of the animal's internal 3-D model of space, possibly using virtual sense data.

Because small animals' representation of space uses a larger brain than an insect brain, and is built from more precise sense data, the parameters of the experimental design may change accordingly. For instance, if a task depends on perceiving collinearity in 3-D space, the testing threshold for non-collinearity may be smaller for larger animals than for insects. The issues of principle in designing experiments are the same.

## 8. Phenomenal Consciousness

If there is a wave in the brain, it has important implications for theories of consciousness. If we assume that consciousness arises from the wave in the brain, then that addresses some of the problems that have beset neural theories of consciousness for many years. Some of these problems are:

1. Which neurons does consciousness arise from? consciousness can persist in the absence of many types of neurons [Penfield & Jasper 1954, Merker 2007], for instance in the cortex or the cerebellum (but notably, not in the thalamus; some lesions in the thalamus permanently interrupt consciousness). This is a problem for any neural theory of consciousness. It is not a problem for a wave theory of consciousness, where consciousness does not depend directly on neurons, only on the wave in the thalamus.
2. The nature of spatial consciousness: Consciousness is like a rather faithful 3-D model of local space; but all known neural representations of space in the brain are complex and highly distorted. How does an undistorted conscious model of space arise from neuron firing? The wave holds a fairly undistorted model of space. If consciousness arises from the wave, we expect it to be an undistorted model of 3-D space – as it is.

---

[7] There is also an evolutionary argument, that animals would not invest large resources in getting high-quality sense data, if their brains could not make the best possible use of it, by building a fast and precise 3-D model of space – close to the best possible model.

3. The unity of consciousness: Many neural processes can go on in parallel in the brain; why then do we have only one serial consciousness? In a wave theory of consciousness, the unity of consciousness can be understood, because there is only one wave in the brain.
4. Why consciousness evolved: If consciousness is a mere side-effect of neural firing, why did it evolve to have the very special form it has? If consciousness arises from the wave, this is not a difficulty. The wave in the brain serves a vital evolutionary purpose – to help an animal represent 3-D space all around it, to plan and control all its movements. The form of the wave has been under intense selection pressure; its form determines the form of consciousness.

If there is a wave in the brain, then there is a compelling case to identify the wave as the origin of consciousness [Worden 1999, 2024c]. The resulting wave theory of consciousness appears promising, because it gives an economical account of the spatial nature of conscious experience – the most important experimental fact about consciousness.

This might be taken as supportive evidence for the wave hypothesis; but because theories of consciousness are controversial, it is best not to rely on it. We can instead go in the opposite direction – from data about consciousness, to the wave hypothesis.

### 8.1 Data about Spatial Cognition from Conscious Experience

Irrespective of theories of the origin of consciousness, what can our consciousness tell us about human spatial cognition? In turn, what can that tell us about the wave hypothesis of spatial cognition?

Consciousness arises from events in the brain. Specifically, the spatial form of our consciousness (which is a rather faithful 3-D model of the space around us) arises from a representation of 3-D space in the brain. In a Bayesian theory of cognition, this is a Bayesian maximum likelihood model, built from sense data (whether that model is stored as neuron firing, or as a wave). This model cannot be much better than our sense data, but it can be very nearly as good.

The form of our conscious awareness gives evidence about the 3-D model of space in the human brain – in particular, about its spatial and temporal resolution. These are both impressive:

- When something changes in our visual field, our conscious experience is updated within about 1/3 second;
- we can consciously detect periodic changes of locations at any frequency less than about 10 Hz;
- Our conscious experience of something near-central in the visual field has a spatial resolution better than one part in 1000;
- We experience a straight line in real space as a straight line in our conscious awareness, with precision better than one part in 100.

Consciousness provides evidence that the brain's internal model of 3-D space is as fast and precise as these numbers. This evidence is independent of any theory of consciousness.

The high precision and speed of conscious experience must arise from comparable underlying high precision and speed in the brain's internal model of space. So the data from our conscious experience can be used to test the computational models of section 6 (neural only, and neural + wave), to help to distinguish between them. It sets a high bar for spatial precision and speed, which poses problems for neural-only models, but poses less serious problems for the wave model.

If there is a wave in the brain which is the origin consciousness, the many different qualia which we can experience at any location of our experienced space imply that the wave excitation has many different degrees of freedom at any wave vector (at any experienced location). An electro-magnetic wave has only two degrees of freedom – another indication that the wave is probably not simply electromagnetic.

## 9. Biophysics of the Wave

In our current state of knowledge, any attempt to understand the biophysics of the wave is a very open-ended theoretical exploration. I only offer a few possible pointers – some promising directions, and some less promising directions. These will change.

### 9.1 The wave probably has very low intensity

As the proposed wave is an essential component of all brains, and needs to be active at every moment of the day (unlike many neural circuits, which are quiescent until needed), its energy consumption could be a serious cost to the animal. We would expect it to have evolved to have the smallest possible energy consumption.

It may therefore have evolved in a direction of decreasing intensity – and the minimum intensity, towards which it evolves, may be very low indeed. Neurons are known to be able to couple to wave excitations (such as light and sound) at very low intensities, down to the single quantum level; or to smells at very low concentrations. The limiting factor is probably not the coupling capability of neurons (or of whichever cells in the brain couple to the wave), but the ability of a signal in the wave to stand out over background noise.

### 9.2 The wave is probably not electromagnetic

There has been interest in electromagnetic fields in the brain, [e.g. Pinotsis, Fridman & Miller 2023, McFadden 2002], so it is natural to ask: could the wave be an electromagnetic wave? It is probably not, for three reasons:

1. **An E-M field or wave is subject to high levels of background noise**: There are known and measured electromagnetic fields in the brain; they are regularly used, for instance in EEG. They are thought to be a consequence of neural electrical activity, by understood physical mechanisms [Nunez & Srivanasan 2006]. Whether or not this E-M field has some computational use, it seems unlikely that its spatial form is determined by the locations of objects around the animal. So for the purpose of storing a 3-D model of local space, the E-M field from neural firing is essentially background noise. This background noise implies that any E-M field or wave used for spatial information storage (as the proposed wave is used) would have to compete with high levels of neural electric noise, and so could not evolve to have low intensity.
2. **An E-M field or wave cannot store information for the required times**: The essential role of the proposed wave is to store spatial information, for periods as long as fractions of a second. However, the E-M fields in the brain are just a passive and instant consequence of neural electrical activity, through Maxwell's equations. They do not store information for fractions of a second. An electromagnetic wave stores information; how it does so is well understood, but it travels at the speed of light, so at typical brain oscillation frequencies of 40Hz, it has a wavelength of 8,000 Km. This would not confine information in a brain, for even 1/40 second. E-M fields are well understood, and they do not do the required information storage function of the wave.
3. **An E-M wave has only two degrees of freedom (polarisations) whereas we can experience many qualia**: This reasoning only applies if the wave is the origin of consciousness; but if the wave exists, it appears highly likely that it is the origin of consciousness, for the reasons described in the previous section. We can experience many different qualia (such as colours) at any location in our conscious experience – that is, at any wave vector of the wave. This would impliy that the wave has more than two degrees of freedom.

**9.3 Investigate Quantum Coherent Effects**

There is increasing interest in quantum coherent effects in biology – for instance in photosynthesis. There is also some evidence of coherent quantum effects in brains [Kersken & Perez 2022]. Quantum effects in brains are reviewed in [Adams & Petruccione 2019].

Quantum coherence can be explored as an evolutionary path to a long-lived wave excitation. Evolution proceeds in small incremental steps, could these steps have been in a direction of increasing quantum coherence times, leading eventually to the times of fractions of a second needed for spatial memory in multi-celled animals?

All of classical dynamics – including classical laws of motion, electromagnetism, locality, fluid dynamics, thermodynamics, and chemical dynamics – can be derived from quantum mechanics by the physics of decoherence [Zurek 2006, Schlosshauer 2010]. This has led to arguments [Tegmark 1999, Koch & Hepp 2006] that because of decoherence, quantum coherence times longer than about $10^{-20}$ seconds cannot occur in a 'warm, wet brain'. The issue is still controversial. Experimentally, these arguments do not apply to a Bose-Einstein Condensate (which acts like a very long-lived quantum state [Feynman, Leighton & Sands 1965]), and we cannot be sure that other long-lived quantum states do not exist.

If there are quantum effects in cognition, they probably exist down to the single-cell level. [Fields & Levin 2021] find from cellular energy considerations that cellular computation must involve quantum coherence.

It may appear that extending quantum coherence times in cells for as long as fractions of a second is very unpromising; but we cannot be certain.

In an analogy, current explorations of coherent quantum effects in cells are like trying to understand muscle action, having only just invented the steam engine. At that time, we might have thought: surely the only way to convert chemical energy to mechanical energy is through heat? Surely the efficiency of energy conversion is limited by the Carnot Cycle and the Second Law of Thermodynamics? This would have implied that, because muscles do not have large internal temperature differences, they must be very inefficient; but muscles are known to be efficient energy converters.

Evolution has invented efficient muscles, and it can exploit extended quantum coherence times, as it does in photosynthesis, bird navigation and possibly olfaction. Condensed matter physics has revealed a large 'zoo' of coherent quantum excitations; we cannot presume that evolution has not found a way to use coherent quantum effects at high temperatures. There is a large space of possibilities to explore.

**9.4 Nuclear and Electronic Spins**

There has been recent interest in the effects of nuclear spins in the brain. Interest has focused on a few elements, notably Lithium, Phosphorus, and Xenon. Two topics have been explored – nuclear spin states with long coherence times, and entanglement of spin states (in the context of possible quantum computation in the brain). The wave proposed in this paper serves as a spatial analogue memory – not as a computing mechanism. So quantum computation and entanglement of spin states are not directly relevant to the wave hypothesis. We are concerned with long-lived nuclear spin states, and their use as a memory, for timescales up to the order of a second. Theoretical considerations of quantum computing are not needed.

A key requirement for the wave in the brain is for it to be insulated from random effects such as thermal fluctuations – so it can work as a memory at low intensities. Nuclear spin systems are well insulated from chemical events in cells, so they might have given evolution a head start in achieving long quantum coherence times. There is an open problem of how living cells can couple to a nuclear spin system; but we should not assume that it is insoluble, unless we can prove so. There are indications of nuclear spin effects in biological matter.

[Fisher 2015] has identified a mechanism whereby the nuclear spins of Phosphorus atoms may have long coherence times in biological matter – for instance (in 'Posner molecules') for times as long as minutes. This mechanism illustrates that nuclear spin states in biological matter may be able retain information for periods as long as would be needed for a spatial working memory.

There is also a hint from anesthesia, which is closely linked to consciousness, and therefore linked to spatial cognition, which is the function of the proposed wave. Xenon, an inert gas, acts as an anesthetic. As Xenon has no chemical reactions, its anesthetic action is not understood. It has been found experimentally in mice [Li et al 2018] that the anesthetic effect of Xenon depends on the isotope of Xenon. Four isotopes have been tested; they differ in their nuclear spins, and in their anesthetic effect. So anesthesia, the interruption of spatial cognition, may depend on nuclear spins.

Similarly there is evidence that the biological effects of Lithium depend on nuclear spin [Sechzer et al 1986].

If the wave in the brain was disabled by nuclear spins, this would interrupt spatial cognition and therefore interrupt consciousness. So it is possible that Xenon acts as an anaesthetic because its nuclear spins disturb a wave which depends on nuclear spins. Further experiments may clarify this.

### 9.6 Bose-Einstein Condensates

A Bose-Einstein Condensate (BEC) [Bose 1924, Einstein 1925] is a remarkable and rare state of matter, which has been studied experimentally and understood theoretically for many years [Pitaevskii & Stringari 2003]. BECs include superconductors, superfluids, and clouds of boson atoms at very low temperatures. A BEC consists of very large numbers of Bose quanta condensed into the same quantum state.

While BECs are not immune to decoherence (there are no long-lived BEC 'Schrodinger Cat' states) [Dalvit et al 2000], certain information in a BEC is immune to decoherence and thermal fluctuations, and persists for very long times. Information persists in second-quantised excitations which have an energy gap from the BEC ground state, and are long-lived. It is a robust experimental fact that a BEC acts like a long-lived quantum state [Feynman, Leighton & Sands 1965].

This would make a BEC a suitable substrate for biological memory – if it can exist in biological matter at the required temperatures. While most known BECs exist well below room temperature, high temperature superconductors have been made close to room temperature, and nature may be able to do better than that.

There are theoretical models of BEC-like states in biological matter, such as [Frohlich 1968]. In Frohlich's model, energy is pumped into the higher of two energy states, so that stimulated emission between the states populates a BEC-like state – as in a laser, which coherently adds photons in a single quantum state. This is of interest for two reasons: first, lasers work at high temperatures; and second, stimulated emission adds quanta which are in phase with existing quanta, and this property might be useful in a wave spatial memory. Frohlich's model has not, as far as I know, been detected in biological matter.

### 9.7 Two areas for search

The considerations in this section have been highly speculative, and are merely indications of areas for possible research. In the search for biophysical mechanisms of a wave, there are two issues to be addressed:

a. The wave needs to persist autonomously for macroscopic timescales
b. Neurons must couple selectively to the wave, both as transmitters and receivers.

Both areas present conceptual challenges; and the two functions may be realised by different cellular mechanisms – different proteins, from the expression of different genes, even in different cell types. For instance, glial cells might support the wave, while neurons couple to it. These are two distinct areas for biophysical investigation.

## 10. Direct Detection of the Wave

The previous section has described reasons why the proposed wave has not yet been detected in the brain. First, it is probably not just electromagnetic, so it would not be picked up by most current detection apparatus; and second, it has probably evolved to have very low intensity, so it will be hard to detect. The lack of direct detection is therefore not strong evidence against the wave hypothesis.

Equally, this does not mean we should give up on trying to detect the wave. Like neutrinos, or like gravitational waves, if we know what to look for, it can eventually be detected. But the lesson from previous discoveries is that we need to know what to look for, in order to build the detectors. We are not yet at that stage; so the direct search for a wave should probably wait for decisive results from the investigations of the previous sections.

## 11. Conclusions

The hypothesis that there is a wave in the brain, representing 3-D space, is a departure from classical neuroscience. If correct, it would mark a fundamental change in how we think of the brain – from being a vast network of neurons, to being an information-rich wave, surrounded by neurons. This change of viewpoint would be as fundamental as Rutherford's nuclear model of the atom.

What are the chances that there is such a wave? Although a wave with the required properties has never been observed, there are good reasons for this. We have not known how to look for it, and we have never examined brains with apparatus that could have detected it.

There are three lines of indirect evidence for the wave – in the central body of the insect brain, in the mammalian thalamus, and in the failure of neural computational models to do 3-D spatial cognition as well as animals do it. Neurons seem to be too imprecise and too slow to do spatial cognition.

Taken together, these three lines of evidence give fairly strong support to the wave hypothesis; [Worden 2024e] estimates the probability of the wave to be robustly greater than 0.4 (and asks for commentary on the result).

So there is a significant chance that there is a wave in the brain; but not a big enough chance to start believing it. This level of uncertainty is unsatisfactory. If there is a wave, we want to know as soon as possible; or if there is not a wave, we need to tackle the serious problems of spatial cognition in a neuron-only brain.

This paper starts to identify the lines of research needed to answer the question – to drive our level of confidence in the wave hypothesis up to near 1.0, or down to near 0.0.

The paper describes research topics involving a wide range of disciplines – from connectomics, genomics and proteomics, to animal behaviour studies, neurophysiology, biophysics, and computational modelling. These are open, green-field research projects – off the well-trodden paths of classical neuroscience, with the possibility of contributing to a revolution in neuroscience.

## References

Adams B. and Petruccione F. (2019) Quantum Effects in the Brain: a Review https://arxiv.org/abs/1910.08423

Bose S. N. (1924) Z. Phys. 26, 178

Chittka, L. (2022) The Mind of a Bee, Princeton University Press, Princeton, NJ

Cowan C.L ; F. Reines; F. B. Harrison; H. W. Kruse; A. D. McGuire (1956). "Detection of the Free Neutrino: a Confirmation". Science. 124 (3212): 103–4.

Dalvit A, Dziarmaga J and Zurek W (2000) Decoherence in Bose-Einstein Condensates: Towards Bigger and Better Schrodinger Cats, arXiv:cond-matt/0001301v2

Dehaene S, Changeux J-P and Naccache L (2011) The Global Neuronal Workspace Model of Conscious Access: From Neuronal Architectures to Clinical Applications, Chapter in Research and Perspectives in Neurosciences

Einstein A. (1925) Sitzber. Kgl. Preuss. Akad. Wiss., 3.

Erman, L. D., Hayes-Roth, F., Lesser, V. R., and Reddy, R. (1980). The HEARSAY-II speech understanding system. Comput. Surv. 12, 213–253.

Feldman, J. (2013). The neural binding problem(s). *Cogn. Neurodyn.* 7, 1–11. doi: 10.1007/s11571-012-9219-8

Feynman R.P., R. B. Leighton and M. Sands (1965) The Feynman Lectures on Physics, Vol III, Addison-Wesley

Fields C. and Levin M. (2021) Metabolic limits on classical information processing by biological cells, BioSystems 209

Fisher,M.P.A.(2015).Quantum cognition: the possibility of processing with nuclear spins in the brain. Ann.Phys. 362

Friston, K. (2003). Learning and inference in the brain. *Neural Netw.* 16, 1325–1352. doi: 10.1016/j.neunet.2003.06.005

Friston K. (2010) The free-energy principle: a unified brain theory? Nature Reviews Neuroscience

Friston K., Kilner, J. & Harrison, L. (2006) A free energy principle for the brain. J. Physiol. Paris 100, 70–87

Frohlich, H. (1968) Long-range coherence and energy storage in biological systems. Int. J. Quantum Chem., 2, 641–649.

Gomez F (2017) The function of the ocelloid and piston in the dinoflagellate Erythropsidinium (Gymnodiniales, Dinophyceae) J. Phycol. 53, 629–641

Hafting, T.; Fyhn, M.; Molden, S.; Moser, M. B.; Moser, E. I. (2005). "Microstructure of a spatial map in the entorhinal cortex". Nature. 436 (7052): 801–806

Hailey A. C. and Krubitzer L. (2019) Not all cortical expansions are the same: the coevolution of the neocortex and the dorsal thalamus in mammals, current opinions in neurobiology 56:78

Halassa, M. M., and Sherman, S. M. (2019). Thalamocortical circuit motifs: a general framework. *Neuron* 103, 762–775. doi: 10.1016/j.neuron.2019.06.005

Hawkins J, Ahmad S. and Cui Y (2017) A Theory of How Columns in the Neocortex Enable Learning the Structure of the World, Frontiers in Neuroscience, doi: 10.3389/fncir.2017.00081

Hebb, D.O. (1949). The Organization of Behavior. New York: Wiley & Sons

Heinze S. et al (2023), the Insect Brain Database, https://insectbraindb.org . Curators of the database are: Berg B.G., Bucher G., el Jundi B, Farnworth M., Gruithuis J., Hartenstein V., Heinze S., Hensgen R., Homberg U., Pfeiffer K., Pfuhl G., Rossler W., Rybak J., Younger M.

Jones, E. G. (2007). The Thalamus, 2nd edition. New York, NY: Cambridge University Press.

Kerskens C and Perez D (2022) Experimental indications of non-classical brain functions, Journal of Physics Communications

Koch, C. & Hepp, K. (2006): Quantum mechanisms in the brain. Nature 440: 611-612.

Knill, D. C., and Pouget, A. (2004). The Bayesian brain: the role of uncertainty in neural coding and computation. *Trends Neurosci.* 27, 712–719. doi: 10.1016/j.tins.2004.10.007

Lee, T. S., and Mumford, D. (2003). Hierarchical Bayesian inference in the visual cortex. *J. Opt. Soc. Am. A Opt. Image Sci. Vis.* 2, 1434–1448. doi: 10.1364/josaa.20.001434

Li N, Lu D, Yang L, Tao H, Xu Y, Wang C, Fu L, Liu H, Chummum Y and Zhang S (2018) Anesthesiology 129 271{277

Llinas, J., and Anthony, R. T. (1993). Blackboard concepts for data fusion applications. *Int. J. Pattern Recognit. Artif. Intell.* 7, 285–308. doi: 10.1142/S0218001493000157

Lynn A, Schneider D and Bruce L (2015) Development of the Avian Dorsal Thalamus: Patterns and Gradients of Neurogenesis, Brain Behav Evol (2015) 86 (2): 94–109.

McFadden, J. (2002) Synchronous firing and its influence on the brain's magnetic field. Journal of Consciousness Studies, 9, 23-50.

McCulloch W.S. & W. Pitts (1943), A logical calculus of ideas immanent in nervous activity, Bulletin of Mathematical Biophysics 5, 115

Merker B (2007) Consciousness without a cerebral cortex: A challenge for neuroscience and medicine, Behavioral & Brain Sciences, 30, 63–134

Moser, M.B., Rowland, D.C., Moser, E.I., 2015. Place cells, grid cells, and memory. Cold Spring Harbor perspectives in biology 7, a021808.

Mueller T (2012) What is the thalamus in zebrafish? Front. Neurosci., 07 May

Mumford, D. (1991). On the computational architecture of the neocortex I: the role of the thalamo-cortical loop. Biol. Cybern. 65, 135–145. doi: 10.1007/BF00202389

Murray, S. O., Olshausen, B. A., and Woods, D. L. (2003). Processing shape, motion and three-dimensional shape-from-motion in the human cortex. *Cereb. Cortex* 13, 508–516. doi: 10.1093/cercor/13.5.508

Nii, P. (1986). The blackboard model of problem solving and the evolution of blackboard architectures. *AI Mag.* 7:38. doi: 10.1609/aimag.v7i2.537

Nunez P, Srinivasan R (2006) Electric Fields of the Brain: The neurophysics of EEG (2nd edn) Oxford Academic

Olshausen, B. A., Anderson, C. H., and Van Essen, D. C. (1993). A neurobiological model of visual attention and invariant pattern recognition based on dynamic routing of information. *J. Neurosci.* 13, 4700–4719.

Olshausen, B. A., Anderson, C. H., and Van Essen, D. C. (1995). A multiscale dynamic routing circuit for forming size- and position-invariant object representations. *J. Comput. Neurosci.* 2, 45–62 doi: 10.1007/BF00962707

Parr T, Pezzulo G, and Friston F (2022) Active Inference: The Free Energy Principle in Mind , Brain and Behaviour, MIT Press, Cambridge, Mass

Parr T, Sajid N, Da Costa L, Mirza M & Friston K (2021) Generative Models for Active Vision, Frontiers in Neurobotics, 15

Pauli, W (1930) "Liebe Radioaktive Damen und Herren" [Dear Radioactive Ladies and Gentlemen].

Penfield, W. & Jasper, H. H. (1954) Epilepsy and the functional anatomy of the human brain. Little, Brown.

Pinotsis, D.A., Fridman, G., and Miller, E.K. (2023) Cytoelectric Coupling: Electric fields sculpt neural activity and "tune" the brain's infrastructure. Progress in Neurobiology, https://doi.org/10.1016/j.pneurobio.2023.102465.

Pitaevskii L. and Stringari S. (2003) Bose-Einstein Condensation, Oxford Science Publications, Oxford, UK

Riesenhuber M. and Poggio T. (1999) Are Cortical Models Really Bound by the "Binding Problem"? Neuron, Vol. 24, 87–93

Rolls,E.T., and Deco,G. (2002). Computational Neuroscience of Vision. Oxford: Oxford University Press.

Schlosshauer M (2010) Decoherence and the quantum to classical transition, Springer

Sechzer J A, Lieberman K W, Alexander G J, Weidman D and Stokes P E (1986) Aberrant Parenting and Delayed

Offspring Development in Rats Exposed to Lithium. Biological Psychiatry 21 1258-1266

Shadlen M. N and Movshon J. A (1999) Synchrony unbound: a critical evaluation of the temporal binding hypothesis; Neuron, vol 24, 67

Sherman, S. M. (2007). The thalamus is more than just a relay. *Curr. Opin. Neurobiol.* 17, 412–422. doi: 10.1016/j.conb.2007.07.003

Sherman, S. M. and Guillery, R. W (2006) Exploring the Thalamus and its role in Cortical Function, MIT Press, Cambridge, Mass

Shigeno S, Andrews P, Ponte G and Fiorito G (2018) Cephalopod Brains: An Overview of Current Knowledge to Facilitate Comparison With Vertebrates, Fronttiers in Physiology Vol. 9

Sjöstedt E, Zhong W, Fagerberg L, Karlsson M, Mitsios N, Adori C, Oksvold P, Edfors F, Limiszewska A, Hikmet F, Huang J, Du Y, Lin L, Dong Z, Yang L, Liu X, Jiang H, Xu X, Wang J, Yang H, Bolund L, Mardinoglu A, Zhang C, von Feilitzen K, Lindskog C, Pontén F, Luo Y, Hökfelt T, Uhlén M, Mulder J. (2020) An atlas of the protein-coding genes in the human, pig, and mouse brain. Science. 2020 367(6482)

Smith R, Friston K, and Whyte C (2022) A step by step Tutorial on Active Inference and its Application to Empirical Data, Journal of Mathematical Psychology, 107, 102632

Spacek J and Lieberman A R (1974) Ultrastructure and three-dimensional organization of synaptic glomeruli in rat somatosensory thalamus. J Anat. Jul; 117(Pt 3): 487–516.

Sprenger J & Hartmann S (2019) Bayesian Philosophy of Science, Oxford

Strausfeld, N. (2011) Arthropod Brains, Belknap, Harvard, Cambridge, Mass.

Tegmark M. (1999) The importance of quantum decoherence in brain processes, Phys. Rev. E61:4194-4206

Treisman, A. (1998). Feature binding, attention and object perception. *Philos. Trans. R. Soc. Lond. B Biol. Sci.* 353, 1295–1306. doi: 10.1098/rstb.1998.0284

Treisman, A., and Gelade, G. (1980). A feature integration theory of attention. *Cogn. Psychol.* 12, 97–136. doi: 10.1016/0010-0285(80)90005-5

Wallis G. and Rolls E.T (1997) Invariant face and object recognition in the visual system, Progress in neurobiology, 51,167

Winding M, Pedigo BD, Barnes CL, Patsolic HG, Park Y, Kazimiers T, Fushiki A, Andrade IV, Khandelwal A, Valdes-Aleman J, Li F, Randel N, Barsotti E, Correia A, Fetter RD, Hartenstein V, Priebe CE, Vogelstein JT, Cardona A, Zlatic M. (2023) The connectome of an insect brain. Science 379(6636): eadd9330.

Von der Malsburg (1995) Binding in Models of Perception and Brain Function, Curr Opin Neurobiol 5(4):520-6

Wallis G and Rolls E (1997) Invariant face and object recognition in the visual system, Prog Neurobiol 51(2):167-94.

Worden, R. P. (1995). An optimal yardstick for cognition. *Psycoloquy* 7:1.

Worden, R.P. (1999) Hybrid Cognition, Journal of Consciousness Studies, 6, No. 1, 1999, pp. 70-90

Worden, R.P. (2010) Why does the Thalamus stay together?, unpublished paper on ResearchGate

Worden, R.P. (2014) the Thalamic Reticular Nucleus: an Anomaly; unpublished paper on ResearchGate

Worden, R. P. (2020a) Is there a wave excitation in the thalamus? arXiv:2006.03420

Worden, R. P. (2020b). An Aggregator model of spatial cognition. arXiv 2011.05853.

Worden R.P, Bennett M and Neascu V (2021) The Thalamus as a Blackboard for Perception and Planning, Front. Behav. Neurosci., 01 March 2021, Sec. Motivation and Reward, https://doi.org/10.3389/fnbeh.2021.633872

Worden R.P. (2024a) The Requirement for Cognition in an Equation, http://arxiv.org/abs/2405.08601

Worden R.P. (2024b) Three-dimensional Spatial Cognition: Bees and Bats, http://arxiv.org/abs/2405.09413

Worden R.P. (2024c) Spatial Cognition: A Wave Hypothesis, http://arxiv.org/abs/2405.10112

Worden R.P. (2024d) The Projective Wave Theory of Consciousness, http://arxiv.org/abs/2405.12071

Worden R. P. (2024e) Assessing the Brain Wave Hypothesis: Call for Commentary: paper to be posted on arXiv

Zurek, W.H. (2006) Decoherence and the transition from quantum to classical —- revisited. Quantum Decoherence, pp 1–31 https ://doi.org/10.1007/978-3-7643-7808-0_1